\documentclass[reprint,superscriptaddress,aps,prb,twocolumn]{revtex4-2}
\usepackage{setspace}
\usepackage{amsmath}
\usepackage{graphicx}
\usepackage{subfig}
\usepackage{ragged2e}
\usepackage{amsfonts}
\usepackage{amssymb}
\usepackage{epstopdf} 
\usepackage{xcolor}
\usepackage{verbatim}
\usepackage{soul}
\usepackage{comment}
\usepackage{placeins}
\usepackage{textcomp}
\usepackage{bm}
\usepackage{hyperref}

\DeclareGraphicsExtensions{.pdf,.eps,.png,.jpg,.mps} 

\usepackage{float}

\begin{document}

\title{Meter-long broadband chirped Bragg gratings for on-chip dispersion control and pulse shaping}

\author{
Zhaoting Geng$^{1,*}$, Yitian Tong$^{1,*}$, Chuchen Zhang$^{2}$, Huajun Tang$^{1}$, Zhenmin Du$^{2}$, Yu Xia$^{1}$, Mingfei Liu$^{1}$, Di Yu$^{1}$, Yuhao Huang$^{1}$, Yaoran Huang$^{1}$, Zheng Li$^{1}$, Tianxiang Dai$^{1}$, Kenneth Kin-Yip Wong$^{1,3}$, Hongwei Chen$^{2,\dagger}$, Chao Xiang$^{1,\dagger}$\\
\vspace{0.5em}
$^1$Department of Electrical and Computer Engineering and State Key Laboratory of Optical Quantum Materials, The University of Hong Kong, Hong Kong, China\\
$^2$Department of Electronic Engineering, Tsinghua University, Beijing 100084, China\\
$^3$Advanced Biomedical Instrumentation Centre, Hong Kong Science Park, Hong Kong, China\\
$^*$These authors contributed equally to this work. \\ 
$^\dagger$Corresponding author: chenhw@tsinghua.edu.cn, cxiang@eee.hku.hk}

\begin{abstract}

Precise on-chip dispersion control is essential for advanced integrated photonic technologies, enabling applications ranging from high‑speed communications and sensing to signal processing and biomedical imaging. However, existing on‑chip dispersion control methods still suffer from substantial loss and a limited dispersion–bandwidth product (DBP) far from application needs. As a result, on‑chip systems continue to rely exclusively on off‑chip dispersion control solutions provided by optical fiber or bulky free‑space optics. To overcome these limitations, we design and fabricate meter-long chirped spiral Bragg gratings (CSBGs) on the ultra-low-loss silicon nitride (SiN) photonic platform for advanced dispersion control. Our device achieves a 10-nanosecond group delay with customizable bandwidths exceeding 10 nanometers within a compact footprint of only 30 $\text {mm} ^2$, surpassing the physical limits of fiber-based grating devices. More importantly, CSBGs can simultaneously possess the characteristics of high stability, low latency, and a large DBP, thanks to the ultra-low-loss SiN platform with a loss of only 0.3 dB/m. Leveraging the precise and stable dispersion profile, we demonstrate high-fidelity pulse shaping and compression of electro-optic frequency combs (EOCs) with a 1-GHz repetition rate centered across the entire reflection bandwidth. The compressed pulse has an on-chip peak (average) power of 21.6 watts (580 milliwatts). Furthermore, we showcase for the first time the application of on-chip pulse-compressed EOC in wavelength-swept coherent anti-Stokes Raman scattering (CARS) microscopy. Our work provides integrated photonics with a long‑sought, scalable, and robust solution for high-performance on‑chip dispersion control, empowering a new generation of on‑chip functionalities.

\end{abstract}
    
\maketitle

Controlling how different wavelengths of light propagate, i.e., the dispersion, is a central task for the operation of modern photonic and optical systems. It underpins a wide range of applications, including high-capacity optical communications \cite{IRN1, IRN2, IRN3}, microwave photonic signal processing \cite{IRN4, IRN5, IRN6}, ultrafast nonlinear optics \cite{IRN7, IRN8, IRN9}, supercontinuum generation \cite{IRN10, IRN11, IRN12}, and so on. In these contexts, optical pulses often accumulate unwanted frequency chirp during propagation, causing temporal broadening that degrades system performance. Conversely, techniques such as chirped pulse amplification \cite{IRN13, IRN14} and time stretch \cite{IRN15, IRN16} require intentionally broadening pulse widths. Achieving the desired flexible pulse manipulation necessitates dispersive elements featured with large, accurate, and manageable group delay dispersion (GDD).
Traditionally, this function relies on bulky components like various diffraction grating pairs or dispersion-compensating fibers (DCFs). Despite their high performance, these dispersion control devices are sensitive to environmental vibrations and occupy significant physical space, hindering the realization of miniaturized, field-deployable dispersion-controlled systems. Although the advent of chirped fiber Bragg gratings (CFBGs) has somewhat alleviated size constraints, their operating bandwidth and achievable dispersion are inherently limited by the processing constraints of grating scribing or writing. Moreover, for emerging on‑chip applications such as integrated pulse shaping \cite{IRN19, IRN20} and signal‑processing \cite{IRN21, IRN22}, the presence of intrinsic coupling losses with external off-chip components can substantially degrade device performance and practical usability.

Therefore, there is a pressing need to transfer dispersion control capabilities onto photonic chips seamlessly with other on-chip elements. On-chip integration has significant potential to deliver compact, scalable, and phase-stable dispersion control solutions by providing a dispersion equivalent to that of kilometer-long optical fibers. However, the transition from fiber-based to chip-based devices has been severely hampered by a critical bottleneck: the excessively high optical propagation loss when large group delay and bandwidth are needed. So far, a significant gap exists between the reported on-chip maximum dispersion-bandwidth product (DBP) and the requirements for practical applications. The typical propagation loss of III-V or silicon‑on‑insulator (SOI) waveguides remains on the order of several dB/cm, which prevents achieving nanosecond-level large group delay without exceeding a practical loss budget. In contrast to III-V and SOI waveguides, silicon nitride (SiN) waveguides feature ultra‑low‑loss characteristics that make them more favorable for achieving larger GDD and broader bandwidth. State-of-the-art high-aspect-ratio SiN waveguides exhibit propagation loss below 0.1 dB/m in the conventional telecom band~\cite{Bauters:11a, Jin2021hertz,Liu2022Ultralow}. Such an extremely low propagation loss allows lightwave to travel meters on the chip while still maintaining strong signal intensity and integrity. Notably, a series of applications requires group delay amounts ranging from several nanoseconds to up to 100 nanoseconds, which correspond to meter-scale on-chip delay lengths. Thus, one of the next major milestones of integrated ultra-low-loss SiN photonics is to fully realize its potential for on‑chip dispersion control, similar to its success in transforming areas such as integrated narrow-linewidth lasers~\cite{li2021reaching,he2024chip} and nonlinear photonics~\cite{liu2020photonic,ji2017ultralowloss,kippenberg2018dissipative,Moss:13}.

In this work, we demonstrate on-chip chirped spiral Bragg gratings (CSBGs) leveraging meter-long ultra-low-loss SiN waveguide structures with high uniformity to achieve a 10-nanosecond on-chip delay with customized delay bandwidth from 1 nm to over 10 nm. Each meter-long spiral structure is packed within a footprint of 30 $\mathrm{mm}^2$. Our CSBGs simultaneously achieve broad bandwidth, low latency, high stability, and a large GDD. 
Such a compact footprint demonstrates the advantages over conventional fiber-based grating devices in scaling to even larger delay and bandwidth.
Based on our CSBG, we demonstrate high-fidelity pulse shaping and compression of an electro-optic frequency comb (EOC) with a 1-GHz repetition rate within the entire reflection bandwidth of the CSBG. 
The compressed pulse with around 13 ps pulse-width has a high peak power of 21.6W, showing the readiness of our approach in various on-chip pulse shaping schemes~\cite{weiner2011ultrafast}.
We further use such stable and wavelength-programmable optical pulses in a wavelength-swept coherent anti-Stokes Raman scattering (CARS) microscopy that operates without mechanical delay compensation. 
In comparison, such performance is nearly unattainable with fiber‑compressed EOCs, whose stability issues fundamentally limit their practicality. Our work thus marks a leap from standalone on-chip dispersion-control components to a functional, enabling engine within a sophisticated photonic system. It also validates the practical utility and performance robustness of our approach over traditional fiber-based approaches, pushing chip-scale dispersion solutions from laboratory concepts into real-world applications in high-speed communications, ultrafast metrology, and advanced biomedical imaging.

\section{Results}

\begin{figure*}[t!]
    \centering
    \includegraphics{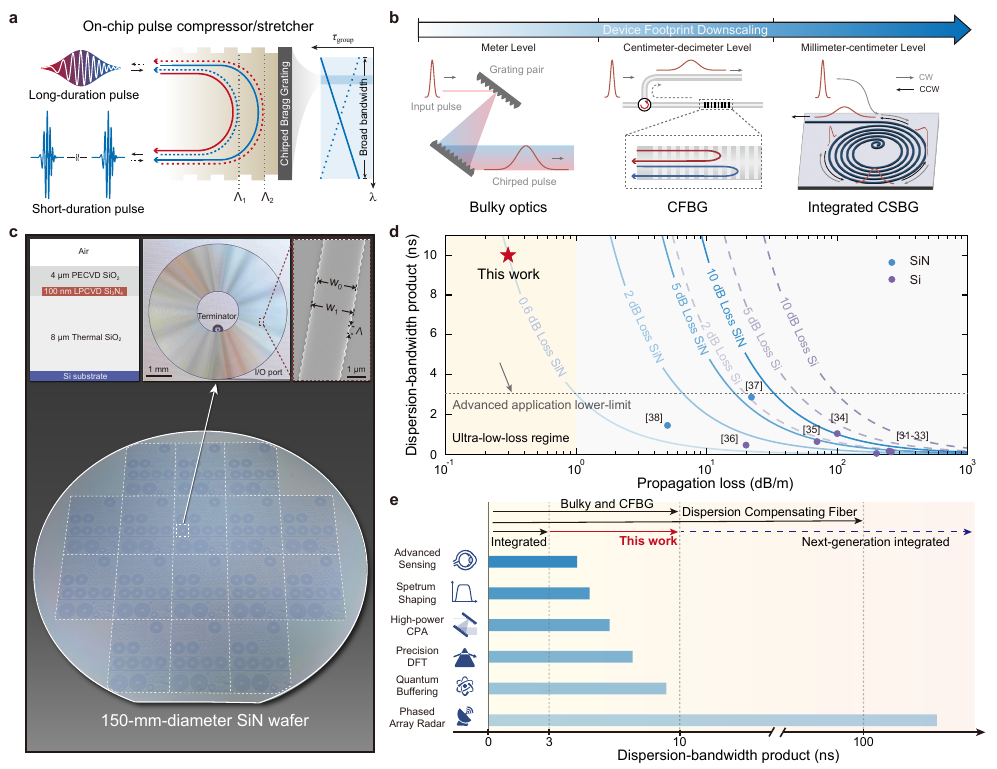}
    \captionsetup{singlelinecheck=off, justification = RaggedRight}
    \caption{\textbf{A summary of devices for dispersion control.}
    \textbf{a}, Concept of an on-chip pulse stretcher/compressor based on chirped Bragg gratings.
    \textbf{b}, Evolution of dispersion control devices: from bulk optics and fiber-based devices to integrated photonics.
    \textbf{c}, A 150-mm-diameter SiN wafer contains hundreds of meter-long chirped spiral gratings. The top three insets (from left to right) show the schematic of the cross-section of the high-aspect-ratio SiN waveguide, the optical microscope picture of a meter-long chirped spiral Bragg grating, and a representative SEM image of a series of fabricated gratings.
    \textbf{d}, Performance benchmark of state-of-the-art on-chip chirped Bragg gratings. The plot illustrates the relationship between the total DBP and the corresponding waveguide propagation loss. The solid lines represent total losses of 0.6 dB (our work), 2 dB, 5 dB, and 10 dB at maximal group delay for the SiN platform, while the dashed lines depict the corresponding values for the Si platform.
    \textbf{e}, The typical requirements of DBP for different applications with the generations of technologies outlined.
    }
    \label{fig:fig1}
\end{figure*}

Fig.\ref{fig:fig1} illustrates the working principle of the on-chip chirped Bragg grating (CBG). By varying the grating period or effective index along a waveguide, a CBG can reflect different spectral components at different positions, leading to spectrally customizable time delay and dispersion. Since the on-chip CBG can offer both positive and negative chirp, it serves as an ideal candidate for pulse compressors and stretchers. One significant advantage of the on-chip CBG is that it transfers the dispersion control function from traditional bulky components to small-sized integrated photonic chips (Fig.\ref{fig:fig1}b). 
However, current on-chip CBGs based on the SOI photonic platform are limited to providing several hundred ps/nm of dispersion. This restricts their use in applications that require high dispersion, such as high-power chirped pulse amplification, dispersive Fourier transform, and phased array radar. The on-chip physical length sets the upper limit on the DBP, and as a result, one cannot achieve both large dispersion and wide bandwidth simultaneously in SOI-waveguide-based devices. Furthermore, high propagation loss results in tilted spectral response if wavelength components need to travel centimeters longer than others before reflecting, practically leading to severe signal distortion. The state-of-the-art ultra-low-loss SiN photonic platform offers the opportunity to break this loss-imposed bottleneck. The ultra-low propagation loss allows lightwave to travel meters on the chip while still maintaining strong signal intensity and integrity. 
Moreover, the moderate index contrast between SiN waveguide core and SiO$_2$ cladding results in reduced phase error than SOI-waveguide-based devices, which is critical in high-uniformity CBGs. We benchmark the performance of current integrated chirped Bragg gratings in terms of propagation loss and the DBP (Fig.\ref{fig:fig1}d)\cite{chenSpiralBraggGrating2015, liuOnChipCirculatorFreeChirped2023, maApodizedSpiralBragg2018, liuOnchipDigitallyTunable2023, sunLargeGroupDelay2021, giuntoniContinuouslyTunableDelay2012,liLargeGroupDelay2023,duSiliconNitrideChirped2020}. Our meter-long CSBG exhibits the lowest propagation loss and the highest DBP among reported works. Furthermore, building on the successful demonstrations of high-Q SiN spiral reference cavity with even longer lengths\cite{jin2021seamless, he2024chip}, we can further extend the length of our CSBGs to more than 10 meters, unlocking new application scenarios for next-generation integrated dispersion control devices requiring over 100-ns-level maximum on-chip delay (Fig.\ref{fig:fig1}e).

\subsection{Architecture and design principles for high-performance CSBGs}

We first design and fabricate our CSBGs on a 150-mm-diameter, CMOS-compatible, ultra-low-loss SiN photonic platform with a 100-nm-thick waveguide layer (Fig.\ref{fig:fig1}c). The detailed fabrication process is provided in the Methods. The width of the waveguide is chosen to be 2.8 \textmu m to ensure single-mode operation. Leveraging this high-aspect-ratio waveguide cross-section design, the waveguide sidewall scattering loss can be largely reduced. The CSBG structure is formed by wrapping the waveguide into an Archimedean spiral and corrugating the waveguide sidewall in a linear chirped period. One port of the Archimedean spiral is used as the reflection-type device’s input/output port, while the other port is connected to a terminator, a small Archimedean spiral waveguide with a gradually narrowing width. The terminator can effectively dissipate light and reduce unexpected reflection. Fig.\ref{fig:fig1}c top-middle inset shows a microscope picture of a meter-long CSBG. The minimal bending radius of the meter-long CSBG is designed to be 800 \textmu m to avoid introducing extra bending loss, meanwhile the outer radius is set to be 2660 \textmu m to ensure the total length is approximately 1 meter. A pitch of 20 \textmu m was selected to minimize crosstalk while maintaining a compact footprint.

\begin{figure*}[t!]
    \centering
    \includegraphics[width=0.9\textwidth]{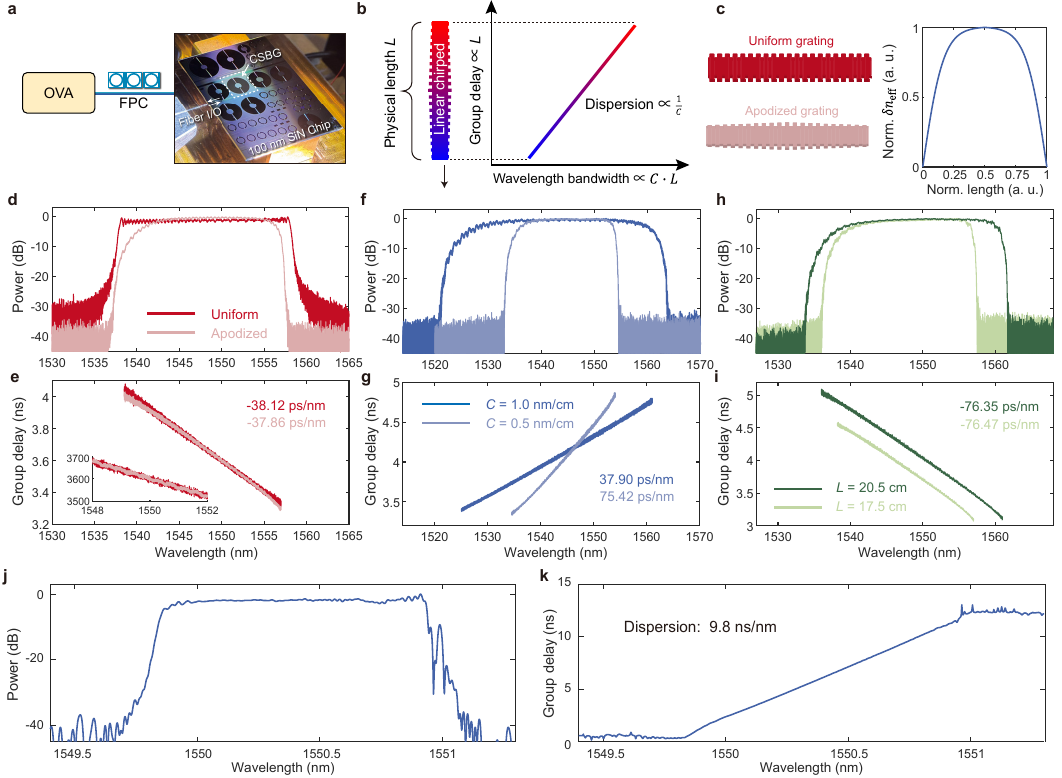}
    \captionsetup{singlelinecheck=off, justification = RaggedRight}
    \caption{\textbf{Dependence of spectral and temporal responses on grating design parameters.} 
    \textbf{a}, Experimental setup for CSBGs characterization. The device is measured using an optical vector analyzer (OVA) and a fiber polarization controller (FPC).
    \textbf{b}, Relationship between grating design parameters and performance.
    \textbf{c}, Schematic of uniform and apodized gratings and apodized function profile.
    \textbf{d,e}, Measured reflection spectra and group delays of uniform and apodized CSBGs.
    \textbf{f,g}, Measured reflection spectra and group delays of two CSBGs with different chirp coefficients \textit{C}.
    \textbf{h,i}, Measured reflection spectra and group delays of two CSBGs with different physical length \textit{L}.
    \textbf{j,k}, Measured reflection spectrum and group delay of a meter-long CSBG with a large dispersion value.
    }
    \label{fig:fig2}
\end{figure*}

In essence, CSBG functions as a distributed reflector with intrinsic spatial dispersion. The period of a CSBG varies linearly along the propagation direction, which changes the Bragg condition at each spatial position. The Bragg condition of CSBGs can be expressed as:
\begin{equation} \label{eq:local_bragg}
    \lambda(z) = 2\bar{n}_{\text{eff}}(z)\Lambda(z)
\end{equation}
where $\lambda$ is the reflection wavelength, $\bar{n}_{\text{eff}}$ is the local average effective refractive index of the waveguide, and $\Lambda$ is the grating period. As the duty cycle $\Gamma$ of the side wall grating is set to be 0.5, the $\bar{n}_{\text{eff}}$ can be simply expressed as:
\begin{equation}
    \bar{n}_{\text{eff}} = (n_{\text{eff,W}} + n_{\text{eff,N}})/2
\end{equation}
where $n_{\text{eff,W}}$ and $n_{\text{eff,N}}$ are the effective refractive indices of the wide and narrow waveguides, corresponding to widths $W_1$ and $W_0$ shown in Fig.\ref{fig:fig1}c top-right inset. Fig.\ref{fig:fig2}b illustrates the basic design concept of CSBGs: The physical length \textit{L} sets the upper limit of the maximum group delay, the DBP, of the CSBG. The chirp coefficient \textit{C}, which describes the degree of variation of the grating period in our case, can be expressed as:
\begin{equation}
    \Lambda(z) = \Lambda_0 + Cz
\end{equation}
where $\Lambda_0$ is the initial period and $\Lambda(z)$ is the period along the chirped grating. By adjusting the value of \textit{L} and the magnitude and sign of \textit{C}, the maximum group delay, reflection bandwidth, and dispersion characteristics (both value and sign) can be easily manipulated. The reflectivity is partly determined by another important factor, the effective refractive index perturbation $\delta n_{\text{eff}}$, defined as:
\begin{equation}
    \delta n_{\text{eff}} = (n_{\text{eff,W}} - n_{\text{eff,N}})/2
\end{equation}
Selecting the appropriate $\delta n_{\text{eff}}$ in different situations is crucial, as an excessively small $\delta n_{\text{eff}}$ results in insufficient reflectivity, while an excessively large $\delta n_{\text{eff}}$ introduces additional scattering loss.

To systematically investigate the influence of the design parameters mentioned above on the characteristics of the CSBG, the fabricated devices consisting of three comparative groups with a single variable altered in each group were experimentally characterized. The experimental setup is illustrated in Fig.\ref{fig:fig2}a. An optical vector analyzer (OVA) with a fiber polarization controller (FPC) is used to measure the CSBG's spectral and temporal responses. 
The first group (Fig.\ref{fig:fig2}d,e) contrasts uniform and apodized gratings. The results indicate that proper apodization significantly suppresses group delay ripples and enhances the reflection roll-off, although with a marginal reduction in the reflection bandwidth and a nonlinear reduction of the group delay at the edge of the bandwidth. In the apodized situation, $\delta n_{\text{eff}}$ is no longer a constant value. It varies along the propagation direction (Fig.\ref{fig:fig2}c), leading to a weak index perturbation at the two ends of the grating to reduce the residual reflections at the grating ends. A hyperbolic tangent profile was adopted as the apodization function for all apodized gratings in this work \cite{ennserOptimizationApodizedLinearly1998}, which can be expressed as follows:
\begin{equation}
    \delta n_{\text{eff}}(z) = \frac{\tanh(\frac{2Az}{L})}{\tanh(A)} (n_{\text{eff,W,max}} - n_{\text{eff,N}})/2
\end{equation}
where $A=3$ is the apodized modulation coefficient.
The second group (Fig.\ref{fig:fig2}f,g) investigates the influence of the chirp coefficient, $C$. Given a fixed grating length, the maximum group delay remains constant. Consequently, a larger chirp coefficient induces a broader variation in the grating period, resulting in an increased spectral bandwidth and a decreased dispersion value. 
The third group (Fig.\ref{fig:fig2}h,i) examines the impact of the grating length, $L$. Since the physical length of the grating sets the upper limit for the maximum group delay, increasing $L$ while keeping $C$ constant directly translates to a wider spectral bandwidth. 
Consequently, for applications demanding the simultaneous achievement of large dispersion and wide bandwidth, longer gratings are indispensable, which need the propagation loss to be kept sufficiently low.

Leveraging the low-loss SiN photonic platform and the design principles outlined above, we demonstrated a meter-long CSBG, with the apodized grating exhibiting a linearly chirped period ranging from 526.5 nm to 526.8 nm. The waveguide parameters are configured with $W_0 = 2.8$ \textmu m and $W_{1,\text{max}} = 2.835$ \textmu m, yielding a $\delta n_{\text{eff,max}} = 1 \times 10^{-4}$. The length of this grating is 1.14 meters. Fig.\ref{fig:fig2}j,k demonstrate the spectral and temporal responses of this meter-long CSBG. It can be seen that the bandwidth of this meter-long CSBG is about 1 nm, and its central wavelength is close to 1550.5 nm. Despite featuring the longest on-chip chirped grating length reported to date, the device exhibits minimal reflectance fluctuations and the lowest spectral tilt compared to previous works \cite{liLargeGroupDelay2023,duSiliconNitrideChirped2020}. Linear fitting of the group delay curve indicates a dispersion of 9.8 ns/nm, representing a record value for integrated photonic systems. Notably, even while maintaining such high dispersion, the device exhibits no observable group delay ripples, rendering it suitable for practical dispersion control applications.

\subsection{Meter-long CSBG for EOC pulse compression}

\begin{figure*}[t!]
    \centering
    \includegraphics{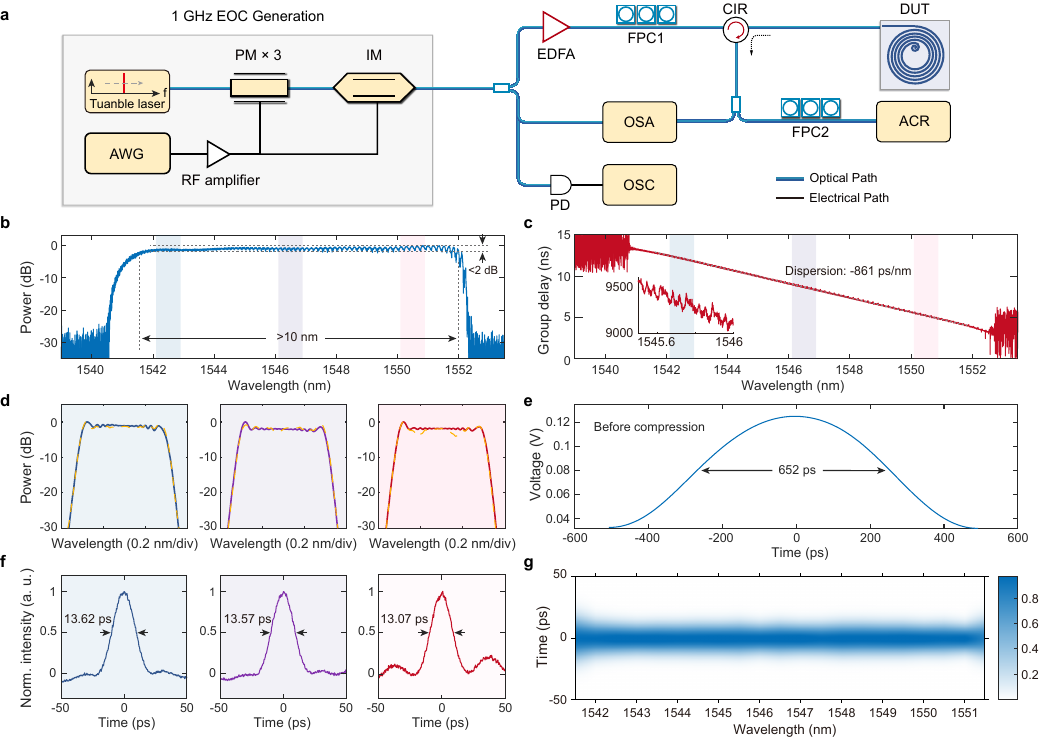}
    \captionsetup{singlelinecheck=off, justification = RaggedRight}
    \caption{\textbf{Characterization of the meter-long CSBG and the on-chip pulse compression results for an EOC with a repetition rate of 1 GHz.}
    \textbf{a}, Measurement setup of the on-chip pulse compression for the 1-GHz EOC. PM, phase modulator; IM, intensity modulator; AWG, arbitrary waveform generator; EDFA, erbium-doped fiber amplifier; FPC, fiber polarization controller; PD, photodetector; OSA, optical spectrum analyzer; OSC, oscilloscope; ACR, autocorrelator; DUT, device under test.
    \textbf{b}, Normalized reflection spectrum of the meter-long CSBG used for pulse compression.
    \textbf{c}, Measured group delay of the meter-long CSBG. The GDD is extracted to be -861 ps/nm through linear fitting. The inset shows the enlarged group delay value to demonstrate the group delay ripples less than 1\%.
    \textbf{d}, Optical spectra of the EOC incident on (blue, purple, and pink solid line) and reflected from (yellow dashed line) the CSBG. The shaded regions (light blue, purple, and pink) highlight the three distinct center wavelengths tuned across the reflection bandwidth of the CSBG.
    \textbf{e}, Direct temporal profile of the input EOC pulse recorded by a high-speed oscilloscope, showing a pulse duration of 652 ps (full-width at half-maximum).
    \textbf{f}, Normalized autocorrelation traces of the compressed pulses corresponding to the center wavelengths of \textbf{e}. The pulse duration is extracted to be 13.62 ps, 13.57 ps, and 13.07 ps, assuming a deconvolution factor of 1.414.
    \textbf{g}, Evolution of the normalized autocorrelation traces across the reflection spectrum of the CSBG.}
    \label{fig:fig3}
\end{figure*}

We now increase the \textit{C} to expand the reflection bandwidth of the meter-long CSBG for dispersion control in broadband systems. Fig.\ref{fig:fig3}b,c show the spectral and temporal responses of the broadband meter-long CSBG. The length of this broadband grating is 1.05 meters. The proposed device exhibits a dispersion control bandwidth of approximately 10 nm, spanning from 1541.5 nm to 1551.5 nm, excluding the spectral edges where significant reflectance fluctuations (\textgreater 2 dB) and nonlinear group delay curves are observed. Linear fitting of the group delay curve indicates a dispersion of $-$861 ps/nm. Notably, the spectral tilt induced by propagation loss is nearly negligible as it is outweighed by minor reflectance fluctuations arising from parasitic random reflections. Such random reflections result from the grating fabrication imperfection, which formed Fabry-Perot resonances between the grating and the waveguide-fiber coupling facet. Beyond the high-quality spectral response, the group delay exhibits high linearity and low ripples, collectively confirming the suitability of the proposed CSBG for practical dispersion control applications.

To further validate the dispersion control capabilities of this CSBG, we demonstrate pulse compression for a 1-GHz EOC with the CSBG. Unlike traditional ultrafast pulse compression and shaping techniques, we chose a low-repetition-rate EOC as the light source for compression, based on three key considerations. First, the EOC exhibits a relatively large initial chirp, making it well-suited for our large-chirp CSBG, which requires specific chirp magnitude and linearity. Second, the spectral bandwidth of the low-repetition-rate EOC is relatively narrow and can be flexibly tuned across the wavelengths, enabling us to evaluate pulse compression performance at various wavelengths within the CSBG reflection bandwidth. Third, such an EOC is easy to synchronize with other light sources, and we can investigate the synchronization stability after pulse compression.

Fig.\ref{fig:fig3}a sketches the setup for EOC generation and CSBG-based pulse compression. The initially generated flat-top EOC featured a 3-dB bandwidth of 0.43 nm, corresponding to approximately 50 comb lines. By fine-tuning the drive voltage of the intensity modulator, the required dispersion value of the compressed pulse can be matched with the meter-long CSBG. After optical amplification, the initial EOC pulse train with an average power of 1 W was coupled into the CSBG, and the compressed reflected pulse train was isolated via a circulator and split for spectral and temporal characterization. The on-chip average power after compression is 580 mW, and the fiber-coupled output power is 280 mW with a single-facet coupling loss of about 2.3 dB. Fig.\ref{fig:fig3}d shows the spectra before and after compression at the randomly selected carrier wavelengths of 1542.5 nm, 1546.5 nm, and 1550.5 nm. The spectral envelope remains well preserved following reflection, although slightly more pronounced ripples were observed at 1550.5 nm, consistent with the device's reflection profile. The initial pulse duration of the EOC was relatively wide, as directly measured by a high-speed photodetector and oscilloscope, with a full-width at half-maximum (FWHM) of 652 ps, as shown in Fig.\ref{fig:fig3}e. After on-chip pulse compression, the pulses from the 1-GHz EOC were effectively compressed to the 10-picosecond-level, measured by an autocorrelator. The corresponding pulse durations of different carrier wavelengths are 13.62 ps, 13.57 ps, and 13.07 ps, respectively, as shown in Fig.\ref{fig:fig3}f. The primary limitation hindering compression to the Fourier-transform-limited pulse-width (8.22 ps) is the higher-order dispersion arising from the nonlinear modulation of the EOC, which cannot be compensated by the linear-chirp characteristics of the CSBG. Further, we tune the seed laser wavelength for the EOC generator at 0.5 nm wavelength steps and record the compressed pulses under different center carrier wavelengths, as shown in Fig.\ref{fig:fig3}g. It is noted that stable pulse compression results can be obtained within the CSBG operating bandwidth, illustrating the broadband dispersion control capability of our meter-long CSBGs.

\subsection{Wavelength-swept CARS microscopy via on-chip pulse compressed EOCs}

\begin{figure*}[t!]
    \centering
    \includegraphics[width=\textwidth]{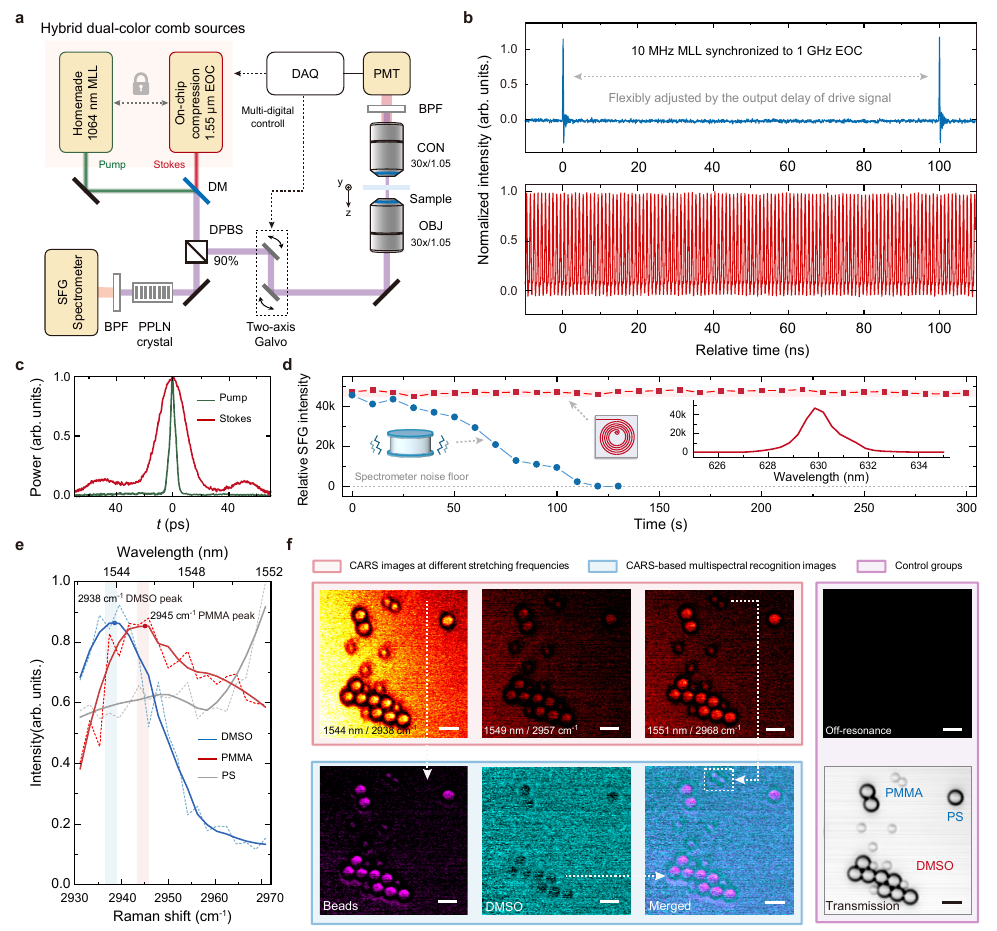}
    \captionsetup{singlelinecheck=off, justification = RaggedRight}
    \caption{\textbf{Wavelength-swept CARS microscopy via the on-chip pulse compressed EOC.}
    \textbf{a}, Scheme of the proposed CARS microscopy.
    \textbf{b}, Two synchronized pulse trains from the hybrid dual-color OFC, which includes a homemade MLL and the proposed 1-GHz EOC based on on-chip compression.
    \textbf{c}, Autocorrelation of the hybrid dual-color OFC. 
    \textbf{d}, Comparison of the synchronization stability between the on-chip compressed pulses and fiber-based compressed pulses via the SFG system. The inset depicts the averaged sum-frequency signal generated via the on-chip compressed pulses.
    \textbf{e}, CARS spectra of DMSO, PMMA, and PS. Dashed line: Data from ten averages of a single point; solid line: Fitted data.
    \textbf{f}, CARS images for three substances at different wavenumbers, multispectral CARS imaging analysis of the three substances via the MCR, and the control groups. DM, dichroic mirror; DPBS, depolarization-beam splitter; SFG, sun-frequency generation; OBJ, objective lens; CON, condenser; BPF, bandpass filter; PMT, photomultiplier tube; DAQ, Data acquisition system. Scale bar: 10 \textmu m.
    }
    \label{fig:CARS}
\end{figure*}

The proposed CSBGs not only perform the pulse compression for EOC over a broad bandwidth, but also ensure high stability of output pulses owing to the integrated SiN platform. This stability is advantageous for precise CSBG reflection spectra in time-domain applications. 
To exemplify this capability, we present the first implementation of wavelength-swept CARS microscopy utilizing on-chip pulse-compressed EOC. CARS microscopy is a label-free, highly sensitive chemical imaging technique that exploits vibrational spectroscopy to produce contrast based on molecular vibrations \cite{CRN1, CRN2, CRN3, CRN4, CRN5}. Central to CARS microscopy is an ultrafast laser system capable of producing two synchronized pulse trains, with one requiring tunable wavelength control to match the vibrational resonance frequencies of target molecules precisely. Typically, pulsed lasers based on mode locking \cite{CRN6} or optical parametric oscillating \cite{CRN7} are employed to generate the pump and Stokes beams. However, due to limitations in wavelength tunability, such light sources for CARS microscopy require complex spatial filtering and pulse shaping, as well as mechanical time-delay calibration \cite{CRN4, CRN8, CRN9}. To address this issue, we replace the conventional MLL (or an optical parametric oscillator) with a synchronized EOC based on on-chip compression, forming a hybrid dual-color optical frequency comb (OFC) for the CARS microscopy, as illustrated in Fig.\ref{fig:CARS}. Leveraging the on-chip pulse-compressed EOC, this CARS microscope eliminates the need for mechanical delay compensation and enables more flexible spectral scanning and higher spectral resolution.

The two synchronized pulse trains maintain a stable 100-fold relationship, as shown in Fig.\ref{fig:CARS}b. The relative phase between them can be rapidly adjusted through digital control of the EOC drive signal, eliminating the need for any additional spatial delay compensation. The autocorrelation results of the output pulses are depicted in Fig.\ref{fig:CARS}c. To evaluate the spatiotemporal synchronization of the pulses, 10\% of the two collimated beams are coupled into an asymmetric sum-frequency-generation (SFG) system \cite{CRN10}. The obtained sum-frequency signal by SFG was served as an indicator of the synchronization stability. For comparative analysis, we evaluated the on-chip compressed pulses against fiber-based compressed pulses within the SFG system, as illustrated in Fig.\ref{fig:CARS}d. Following the optimization of the phase delay, the on-chip pulse compressed EOC achieved effective spatiotemporal synchronization with the pulse output of the 1-\textmu m MLL. A relatively stable sum-frequency signal is observed over a 300-second measurement period, as indicated by the red dots in Fig.\ref{fig:CARS}d. In contrast, due to environmental disturbances in the 1-km fiber used for compression, the compressed pulse experienced temporal jitter, impairing precise time synchronization. Consequently, the amplitude of the sum-frequency signal diminishes to the system's noise floor within approximately 100 seconds. This demonstrates that our on-chip pulse compression not only enables broadband spectral pulse compression but also maintains high stability.

As depicted in Fig. \ref{fig:fig3}g, the 1-meter-long CSBG effectively compresses EOC into stable picosecond pulses within a 10-nm spectral bandwidth. This capability enables the utilization of synchronized EOC as a wavelength-swept pulse source for spectral scanning in CARS microscopy to achieve a high spectral resolution just by programmatically controlling the CW laser. Thus, the two excitation combs constitute a CARS detection region from 2960 $cm^{-1}$ to 2960 $cm^{-1}$. We chose two types of plastics, polystyrene (PS) and polymethyl methacrylate (PMMA), along with dimethyl sulfoxide (DMSO) as the samples. A wavelength scanning procedure was conducted at 0.5 nm intervals for the seed laser of the EOC. At each wavelength, the CARS signal amplitude was recorded and subsequently normalized, as presented in Fig.\ref{fig:CARS}e. DMSO exhibits a pronounced C-H Raman band at approximately 2938 $cm^{-1}$ Raman shift, while PMMA displays a distinct Raman band near 2945 $cm^{-1}$. Although no Raman peak for PS is observed within the measured spectral range, PS demonstrates higher CARS signal intensity at higher wavenumbers relative to the other two substances. The reflection bandwidth of the CSBG limited the detection of the additional Raman peak of DMSO, thereby constraining a comprehensive evaluation of the Raman spectral resolution. Nonetheless, an approximate spectral resolution of around 10 $cm^{-1}$ can be inferred from the FWHM spectra of the two excitation combs, which surpasses that of many conventional CARS microscopes.

By employing the rapid wavelength sweeping, we can differentiate the three aforementioned substances through multispectral CARS imaging, as demonstrated in Fig.\ref{fig:CARS}f. The sample is the DMSO suspension containing two types of plastic beads. The multispectral CARS images corresponding to different Raman shifts are presented within the blue box in Fig.\ref{fig:CARS}f. At the wavenumber of 2938 $cm^{-1}$, the background intensity of DMSO is more pronounced than that of the two microplastic spheres. In contrast, as the wavenumber increases and the spectral position moves away from the strong Raman peak of the DMSO, the imaging contrast of the microplastic beads becomes more prominent. Multivariate curve resolution (MCR) \cite{CRN11, CRN12} was applied to multispectral CARS imaging analysis of the three samples, as shown in the orange box in Fig.\ref{fig:CARS}f. MCR analysis enabled the mapping of the major components within the detection area based on their spectral features in the overlapped C-H region. Consequently, the two types of microplastic beads were effectively distinguished from the DMSO background. The control group comprises the non-resonant background image acquired at corresponding locations, as well as the bright-field image obtained using a 1.0-\textmu m MLL source, as illustrated within the purple box in Fig.\ref{fig:CARS}f.

\section{Discussion}

In summary, this work demonstrates a chip-based device for dispersion control on an ultra-low-loss SiN platform. We transitioned this technology from a theoretical concept to a functional tool by compressing an EOC for wavelength-swept CARS microscopy. This first-of-its-kind application proves that the device is robust and ready for real-world bio-imaging and spectroscopy.

Our CSBGs show clear advantages over commercial fiber solutions in several aspects. To equal the large dispersion provided by our integrated device, one would need nearly 100 km of DCF. This massive length would introduce over 50 dB loss and microsecond-level latency. In contrast, our device maintains only 0.6 dB on-chip insertion loss, 4.6 dB round-trip coupling loss, and signal latency of around tens of nanoseconds, which is critical for high-speed networks. 
Furthermore, the device exhibits excellent power handling capabilities compared to other integrated solutions such as III-V or SOI waveguides. 

Out of many real-world applications, CARS microscopy is one example that requires precise chirp control, broad operational bandwidth, high stability against delay jitter, and flexible adjustability.
Meeting all these requirements demonstrates the robustness in performance and integration maturity of our devices and the readiness for broader deployment in advanced photonic systems.

Looking forward, the ultra-low loss of our platform implies no fundamental limits to the delay length, except being bounded by the mask reticle or wafer sizes. We can readily extend the length of our CSBG to more than 10 meters to enable even greater dispersion control~\cite{jin2021seamless}. Moreover, the full CMOS compatibility opens the door to 3D heterogeneous integration with other on-chip elements \cite{xiang20233d}. For instance, by integrating these devices with on-chip EOCs, femtosecond amplifiers, or soliton microcombs, we can realize fully self-contained, high-performance femtosecond pulse sources \cite{yuIntegratedFemtosecondPulse2022,gaafar2024femtosecond,xiang2021laser}. Ultimately, this work establishes a scalable foundation for integrated dispersion control, substituting a key missing piece for on-chip compact ultrafast spectroscopy, high-capacity optical interconnects, and portable biomedical imaging.

\bibliographystyle{unsrt}
\bibliography{main}

\vspace{10 mm}
\noindent\textbf{Methods}

\noindent \textbf{Device fabrication} A 100-nm-thick stoichiometric silicon nitride film was deposited using low-pressure chemical vapor deposition (LPCVD) on a 150-mm-diameter silicon substrate with an 8-\textmu m-thick thermal oxide. The CSBG waveguide was patterned by the standard deep ultraviolet stepper (DUV) lithography and formed by anisotropically inductively coupled plasma dry etching in a CHF${_3}$/O${_2}$ chemistry. The remaining photoresist and the resulting byproducts of the etch were then ashed by oxygen plasma and completely stripped by soaking in a hot N-methyl-2-pyrrolidone solution and piranha solution. Subsequently, the patterned waveguide was annealed at 1150℃ in a nitrogen atmosphere for 2 hours. Then a 4-\textmu m-thick silicon dioxide upper cladding was deposited using plasma-enhanced chemical vapor deposition (PECVD) with tetraethoxysilane (TEOS) as a precursor, followed by a final anneal at 1150℃ in a nitrogen atmosphere for 16 hours. The fabrication process flow is illustrated in Supplementary Fig. S1.

\medskip

\noindent \textbf{Experimental setup for EOC generation and CSBG-based pulse compression} The optical pulse train is initiated within the 1-GHz EOC module (dashed enclosure). The EOC was generated using a CW tunable laser (CTL 1550, TOPTICA) as the coherent seed. The CW laser was routed through a cascade of three phase modulators followed by an intensity modulator. The phase modulators were employed to broaden the optical spectrum, while the intensity modulator was used to carve the temporal pulse profile. All modulators were driven synchronously by 1-GHz radio-frequency signals synthesized by an arbitrary waveform generator (M8199, Keysight), resulting in a stable optical pulse train with a 1-GHz repetition rate. The frequency and time-domain characteristics of the generated EOC were characterized using an optical spectrum analyzer (AQ6370E, Yokogawa) and a real-time oscilloscope (SDS7304A, Siglent), respectively. The initial EOC is amplified before the dispersion control of the CSBG by an erbium-doped fiber amplifier to compensate for the coupling loss and chip propagation loss. After on-chip pulse compression, the pulses from the EOC were effectively compressed to the picosecond level, measured by an autocorrelator (PulseCheck, APE). A fiber polarization controller was used to control the polarization state of light entering the 1-m-long CSBG. The reflected pulse train was isolated by the circulator and split to simultaneously measure the compressed pulse duration and record the optical spectrum. 

\medskip

\noindent \textbf{Wavelength-swept CARS microscope} The CASRS microscope involves two synchronized laser pulse trains: a pump train and a Stokes train. When the frequency difference between the pump and Stokes pulse trains matches a specific molecular vibrational frequency, a coherent anti-Stokes signal is generated at a higher frequency than the pump beam. This anti-Stokes signal is detected to produce a contrast image highlighting specific molecular species within the sample. In our system, the pump pulse train was provided by a homemade mode-locked Yb-fiber laser at 1064 nm with an 8 nm FWHM, and the Stokes pulse train by the on-chip pulse-compressed EOC at 1550 nm with a 0.43 nm FWHM. The repetition rate of the 1.0-\textmu m MLL is matched to 10 MHz. Thus, the synchronized method leverages the MLL as a master oscillator to provide a common reference for the EOC, ensuring both combs’ phase-locking in the time domain. The autocorrelation of the hybrid dual-color OFC is shown in the inset of Fig.\ref{fig:CARS}a. More details about the sources and the temporal synchronization of the hybrid dual-color OFC are discussed in the Supplementary Materials. The two beams of the hybrid dual-color optical frequency comb were coupled into free space using collimators. To achieve optimal spatial overlap, both beams were transmitted through a dedicated four-focal-length (4f) relay system designed for specific wavelengths, resulting in an output beam waist diameter of 3.6 mm. Then the two beams were collinearly combined via a dichroic mirror (DM). After passing through a depolarization-beam splitter (DPBS), 90\% of the two collimated beams were directed into a scanning unit equipped with a pair of Galvo mirrors, conjugated to the back aperture of the objective lens (UPLSAPO30XSIR, EVIDENT) in the microscope by a 4f system, to facilitate laser scanning imaging. The two OFCs delivered output powers of 200 mW@ 1 \textmu m and 500 mW@1.55 \textmu m, respectively, with corresponding average powers incident on the samples of 70 mW and 100 mW. After the light collection by the same numerical aperture condenser, the anti-Stokes signal was filtered out optically and electronically, and finally input into a photomultiplier tube (PMT). The sum-frequency signal generated by the SFG system was optically filtered out and measured using a visible light spectrometer (USB2000+, Ocean Optics). The phase delay of the hybrid dual-color OFC can be determined by the strength of the sum-frequency signal in the SFG system. Alternatively, a field-programmable gate array can be employed to provide real-time feedback on phase shift adjustments during the scanning process, enabling dynamic phase control. 

\medskip

\noindent \textbf{Preparation of the samples for CARS microscopy} For the CARS spectral analysis, pure samples of PS and PMMA were purchased as 1 mm-thick plastic plates. These samples were sectioned into 3 cm × 3 cm squares and subjected to point scanning for spectral measurements. For the CARS images, PMMA beads (5 \textmu m diameter) and PS beads (10 \textmu m diameter) were suspended in ultrapure water at a concentration of 25 mg/mL. Equal volumes of each suspension were combined and subsequently diluted 1,000-fold with ultrapure water, followed by thorough agitation. A 1 mL aliquot of the diluted mixture was subsequently spin-coated onto a glass slide and allowed to air dry. Afterwards, an appropriate volume of DMSO solution diluted with water at a 1:4 ratio was added, and a coverslip was carefully positioned atop the sample. The edges were sealed with adhesive to ensure secure fixation during subsequent analysis.

\medskip

\medskip

\vspace{5 mm}

\noindent \textbf{Data Availability}
The data that support the plots within this paper and other findings of this study are available from the corresponding author upon reasonable request.

\vspace{2 mm}

\noindent \textbf{Code Availability}
The codes that support the findings of this study are available from the corresponding author upon reasonable request.

\vspace{2 mm}

\noindent \textbf{Acknowledgments}
We thank the funding support from the National Key R\&D Program of China (2024YFA1409300), the Research Grants Council of Hong Kong (C7143-25Y, N\_HKU774\_25, T46-705/23-R, STG3/E-704/23-N, STG3/E-104/25-N), the Innovation and Technology Commission of Hong Kong (GHP/230/22GD), the National Natural Science Foundation of China (6232290014), the Guangdong Provincial Quantum Science Strategic Initiative (GDZX2304004, GDZX2404002), and the Croucher Foundation. A portion of this work was performed in the HKUST Nanosystem Fabrication Facility. 
We thank Dr. Ruixuan Chen for the help in figure preparation.

\vspace{2 mm}

\noindent\textbf{Author Contributions} 
Z.G. Y.T. and C.X. conceived the idea.
Z.G. and Z.D. led the design of the CSBG.
Z.G. fabricated the devices with the assistance of Y.H.,Y.H. and Z.L.
Z.G., Z.D. and C.Z. characterized the spectral and temporal responses of the device.
Y.T. and Z.G. conducted the EOC pulse compression experiment, with contributions from Y.X., M.L., D.Y. and H.T.
Y.T. and H.T. performed the wavelength-swept CARS microscopy experiment.
Z.G. and Y.T. prepared the figures and wrote the manuscript with input from all authors.
C.X., H.C. and K.W. supervised the project.
\vspace{3 mm}

\noindent \textbf{Competing Interests} The authors declare that they have no competing interests.

\vspace{3 mm}

\noindent \textbf{Author Information} Correspondence and requests for materials should be addressed to H.C and C.X. (chenhw@tsinghua.edu.cn and cxiang@eee.hku.hk).

\end{document}